\documentstyle[11pt]{article}

\title{Magnetic and spin evolution of isolated neutron stars
	with the crustal magnetic field}

\author {V. Urpin$^{1,2)}$ and D. Konenkov$^{2)}$\\
	 $^{1)}$ Department of Mathematics, \\
	 University of Newcastle,
\\
	 Newcastle upon Tyne NE1 7RU, UK \\
        {\it e-mail: Vadim.Urpin@newcastle.ac.uk}\\
	 $^{2)}$ A.F.Ioffe Institute of Physics and Technology,
	 194021 St.Petersburg, Russia \\
         {\it e-mail: dyk@astro.ioffe.rssi.ru}}

\date{2 July 1997}

\begin{document}

\maketitle

\begin{abstract}

We consider the magnetic and spin evolution of isolated neutron
stars assuming that the magnetic field is initially confined to 
the crust. The evolution of the crustal field is determined by
the conductive properties of the crust which, in its turn,
depend on the thermal history of the neutron star. Due to this 
fact, a study of the magnetic field decay may be a powerful diagnostic 
of the properties of matter in the core where the density is above the 
nuclear density. We treat the evolution of neutron stars for different
possible equations of state and cooling scenarios
(standard and, so called, accelerated cooling). 

The spin evolution is strongly influenced by the behaviour of the 
magnetic field. Assuming that the spin-down rate of the neutron star 
is determined by the magnetodipole radiation, we calculated the 
evolutionary tracks of isolated pulsars in the $B - \tau$ and $B-P$ 
planes, where $B$ and $P$ are the magnetic field and period, 
respectively, and $\tau$ is the spin-down age. The calculated
tracks are compared with observational data on the magnetic field
and period of pulsars. This comparison allows to infer the most 
suitable equations of state of nuclear matter and cooling model 
and to determine the range of parameters of the original magnetic 
configurations of pulsars.

\end{abstract}

\newpage
\section{Introduction}

Our knowledge of the properties of matter in neutron star
cores is a subject of many uncertainties. The core may 
consist of the normal nuclear matter with "standard" properties,
some components of this matter may be in superfluid or
superconductive state, and the presence of exotic phases 
(free quarks, pion condensate etc.) cannot be excluded
{\it a priori} as well. The most direct way to study the
nature of a superdense matter is a comparison of surface
thermal radiation measurements with predictions of neutron
star cooling models (see, for example, Van Riper 1991,
Nomoto \& Tsuruta 1987). Unfortunately, observations of 
thermal emission from neutron stars are few, and available
observational data don't allow to infer
the properties of the neutron star interiors (see, e.g.,
\"Ogelman 1994). 

Another way to study neutron star cores is associated
with the evolution of the magnetic field. A surface
strength of the magnetic field is estimated for the majority 
of 700 known pulsars of different ages. The evolution
of the field is mainly determined by the conductive properties
of plasma which, in its turn, depend on the temperature. The
thermal evolution is strongly influenced by the state of matter 
in the central region of the star. The link between the
magnetic evolution and properties of the core matter provides
generally one more window into the neutron star interior.  

The evolution of the magnetic field, however, depends on its
configuration which is unknown since the origin of the field
is not clear until now. The field could have been amplified
from the weak field of the progenitor star due to the magnetic
flux conservation. In this case, the magnetic field lines occupy
probably a substantial fraction of the neutron star volume and passes
through the core. It is also possible, however, that the magnetic field 
is confined to relatively not very deep layers. Such a crustal
magnetic configuration can be generated, for example, by the 
thermomagnetic instability during the first years of the neutron 
star life (Urpin, Levshakov \& Yakovlev 1986) or by convection which 
arises in first 10-20 seconds after the neutron star is born (Thompson \&
Duncan 1993). The electric currents maintaining
crustal magnetic fields are anchored in the crust but, due to
a dependence of the crustal conductivity on the temperature,
the evolution of such configurations is also influenced by the
properties of matter in the core. This effect has been
first argued by Urpin \& Van Riper (1993).

The present paper considers in more detail the influence of the
internal structure of a neutron star on the decay of crustal 
magnetic fields. The crustal field decay has been a subject
of study for a number of papers (see, e.g., Sang \& Chanmugam 1987,
1990, Urpin \& Muslimov 1992). These papers have been addressed, 
however, to the magnetic evolution of the particular model of 
a neutron star and did not examine the effect of equation
of state and cooling scenario on the decay. In the present study, 
we treat the behaviour of the magnetic field for the neutron star 
models with different equations of state (BPS, FP, PS) and
different thermal histories. The representatives of cooling
scenarios are the so called "standard" cooling (associated with
the standard neutrino emissivities; see, e.g., Van Riper 1991) and
"accelerated" cooling which can be caused by the enhanced neutrino
emission due to either direct URCA processes (Lattimer et al. 1991) 
or the presence of a pion-condensate or free quarks in the core. 
It turns out, that the magnetic evolution is strongly sensitive
to details of the internal structure, and it provides a hope
to discriminate between theoretical models by a comparison with 
observational data on pulsar magnetic fields. 

Our knowledge of the magnetic field strength of neutron stars
comes mainly from radio pulsars with measured spin-down rates.
The real ages of these pulsars are usually unknown but one can
determine from observational data the so called "spin-down" age,
$\tau = P / 2 \dot{P}$, where $P$ is the spin period of the neutron
star. Therefore, for a comparison with observational data, it is 
more convenient to analyse a dependence of the magnetic 
field on the spin-down age but not on the real age. Assuming 
that the spin-down torque on a pulsar is determined by its magnetic 
dipole radiation, we calculate the magnetic and spin evolution of 
neutron stars and plot the evolutionary tracks in the $B - \tau$ and 
$B-P$ diagrams. A comparison of these tracks with the observable 
distribution of pulsars in the $B - \tau$ and $B-P$ planes allows 
to determine the most suitable parameters of the initial magnetic 
configurations of neutron stars.

The paper is organized as follows. The main equations governing 
the magnetic and spin evolution of neutron stars with the crustal 
magnetic configuration are presented in Section 2. The numerical
results are described in Section 3. In Section 4, we 
compare the theoretical decay curves with the observable distribution
of pulsars and determine the parameters of their magnetic 
configurations. The summary of our results is given in Section 5.

\section{Basic equations}

We assume that the magnetic field of the neutron star has been
generated in the crust by some unspecified mechanism during
or shortly after a neutron star formation. The main fraction of
the crust volume quickly ($< 1$ yr) solidifies, thus one can 
neglect the effect of fluid motions. In the solid crust, the
evolution of the magnetic field, $\vec{B}$, is governed by the
induction equation  
$$
\frac {\partial \vec{B}} {\partial t} =
  -\frac{c^{2}}{4 \pi} \nabla \times 
	\left( \frac{1}{\sigma} \nabla \times \vec{B} \right),  \eqno(1)
$$
where $\sigma$ is the electrical conductivity. We neglect in this 
equation the anisotropy of the conductivity that is justified if
$\omega_{B} \tau_{e} < 1$ where $\omega_{B}$ is the gyrofrequency of 
Fermi electrons and $\tau_{e}$ is their relaxation time. 
We consider the evolution of a dipole magnetic field alone. In
this case, $\vec{B} = \nabla \times \vec{A}$ and the vector potential, 
$\vec{A}$, may be written in the form $\vec{A}=(0,0,A_{\varphi})$;
$A_{\varphi}=s(r,t) \sin{\theta} /r$ where $r$, $\theta$ and
$\varphi$ are the spherical coordinates. The function $s(r,t)$
satisfies the scalar equation
$$
\frac{4\pi\sigma}{c^2} \frac{{\partial} s}{{\partial} t}=
   \frac{{\partial}^2 s}{{\partial}r^2}-\frac{2s}{r^2}. \eqno(2)
$$
We impose the standard boundary condition for a dipole field
$R \partial s/\partial r + s =0$ at the surface $r=R$. For the
crustal field, $s$ should vanish in deep layers. At the late
evolutionary stage, the field can diffuse in relatively deep
layers of the crust and even reach the crust-core boundary. 
We assume that the core matter is in the superconductive
state and the magnetic field does not penetrate into the core.
Note that the dipole configuration is chosen only for the sake 
of conveniency, and higher multipoles can be treated in a same way. 
Their evolution is qualitatively similar to the evolution
of the dipole field.

The field decay is determined by the conductive properties of
the crust which, in its turn, depend on the mechanism of
scattering of electrons. Likely, scattering on phonons and
impurities are most important in the neutron star crust; which
mechanism dominates depends on the density $\rho$ and temperature
$T$. The total conductivity of the crust is given by 
$$
\sigma = \left( \frac{1}{\sigma_{ph}} + \frac{1}{\sigma_{imp}}
\right)^{-1} \;. \eqno(3)
$$
The phonon conductivity, $\sigma_{ph}$, is proportional to $T^{-1}$
when $T$ is above the Debye temperature and to $T^{-2}$ for
lower $T$ (Yakovlev \& Urpin 1980). The impurity conductivity, 
$\sigma_{imp}$, is practically independent of $T$ and its
magnitude is determined by the impurity parameter $Q$: 
$$
Q =\frac{1}{n} \sum_{n'} n' (Z-Z')^{2} \;, \eqno(4)
$$
where $n'$ is the number density of an interloper specie of charge
$Z'$; $n$ and $Z$ are the number density and charge of the dominant
background ion specie; summation is over all species. In the present 
computations, we use the improved numerical results for $\sigma_{ph}$   
obtained by Itoh, Hayashi \& Kohyama (1993) for the so called equilibrium
chemical composition and the analytical expression for $\sigma_{imp}$
derived by Yakovlev \& Urpin (1980).

The conductivity depends on the crustal temperature which is
determined by the thermal evolution of the neutron star. In its turn,
the thermal history is determined by the properties of matter in
the core where the density is above the nuclear density. Due to
this, the magnetic evolution of the neutron star with the crustal
magnetic field is influenced by the state of interiors and,
in principle, can provide information on the properties of
superdense matter. 

With the exception of a short initial phase ($1-10^{3}$ yr
depending on the model) a temperature non-uniformity is
small over the main fraction of the crust volume. Since the
time scale of the decay of the magnetic field is typically longer
than this initial phase, we assume the crust to be isothermal
with $T=T_{c}$ where $T_{c}$ is the temperature at the bottom of 
the crust.

The calculations we present here are based on cooling scenarios
considered by Van Riper (1991) for different neutron star models.
For our purposes, we have selected the cooling curves for models
with a baryon mass of $1.4 M_{\odot}$ constructed with the 
equations of state of Pandharipande and Smith (1975; hereafter
PS), Friedman and Pandharipande (1981; hereafter FP), and Baym,
Pethick and Sutherland (1971; hereafter BPS). The BPS model is
representative of soft equations of state, which result in high
central densities and small mass of the crust. The PS model was
chosen as a stiff model with a low central density and a massive 
crust. The FP equation of state is a representative intermediate
model.The stiffer the equation of state,
the larger are the radius and crustal thickness for a given
neutron star mass. Thus, the radii are  $7.35$, $10.61$ and
$15.98$ km for the BPS, FP and PS models, respectively; the
corresponding crustal thicknesses are $\approx 310$, $940$
and $4200$ m; we assume the crust bottom to be located at the
density $2 \times 10^{14}$ g/cm$^{3}$. Our choice of models is 
not exhaustive but it illustrates well the role of the equation of
state in the magnetic evolution of neutron stars.

For each equation of state, we consider two substantially different cooling
scenarios which represent the slowest ("standard") and most rapid 
("accelerated") cooling models among the results given by Van Riper 
(1991). The standard cooling model corresponds to a star with normal 
$npe$-matter in the core and with standard neutrino emissivities. The 
accelerated cooling represents the models with enhanced neutrino emissions 
in the core either due to the direct Urca processes in the $npe$-matter 
or due to the presence of quarks or pion condensate. Note that the cooling 
scenario is not generally independent of the equation of state. For 
instance, enhanced neutrino emission may result from exotic constituents 
of matter but these constituents can be formed at a very high density 
which can be reached only for soft equations of state. Due to this, the 
models with a stiff equation of state and accelerated cooling or with 
a soft equation of state and standard cooling look questionable. However, 
our knowledge of the phase transition from the normal nuclear matter to 
the exotic one is at best uncertain therefore we present here the
results for all combinations of cooling scenarios and choosen equations
of state.

The standard models have a high surface temperature ($\geq 10^{6}$ K) 
during the first $\sim (3-5) \times 10^{5}$ yr. For the accelerated
cooling models, the surface temperature declines abruptly
at $\sim 1-10^{3}$ yr (depending on the equation of state) and after that
the star cools down relatively slowly. For conveniency, we plott
in Fig.1 the dependence of the internal temperature on the age
for different neutron star models. Note that cooling models given by Van 
Riper (1991) end when $T_{s}= 3 \times 10^{4}$ K, $T_{s}$ is the surface 
temperature. The results of additional cooling calculations for ages 
up to $10^{9}$ yr (as well as the density profiles for different models) 
have been kindly provided us by Prof K.Van Riper. These calculations    
involve balancing the heat capacity of degenerate fermions against 
surface radiation and extrapolation of the smooth $T_{m}/T_{s}(T_{m})$
from $\approx 5$ to 1. However, at these low temperatures ($T < 10^{6}$ K), 
$\sigma$ is dominated by impurity scattering in the region where
the currents maintaining the magnetic field are concentrated, 
and the field decay is thus insensitive to the details of the cooling. 
Due to this, also different reheating mechanisms which can make the 
evolution at late times non-linear cannot affect appreciably the 
behaviour of the magnetic field.

Equation (1) with the corresponding expressions for conductivities
determines the dependence of the neutron star magnetic field on
the age. However, our knowledge of the surface field strength
and its behavior with time comes mainly from radio pulsars with 
measured spin-down rates. For the most of these pulsars, the real
age is unknown and observational data provides information only
on the so called "spin-down" age, $\tau = P/2\dot{P}$, where
$P$ and $\dot{P}$ are the spin period and spin-down rate,
respectively. Therefore, for a comparison with observational data, it 
is more convenient to consider the dependence of B on $\tau$ and $P$
rather than on $t$. With the assumption that the spin-down
torque on the pulsar is determined by its magnetodipole radiation
(Ostriker \& Gunn 1969), the spin period and spin-down rate are 
related to the surface field strength at the magnetic equator, $B_{s}$, 
by
$$
P \dot{P} = \alpha B_{s}^{2} \;, \eqno(5)
$$
where $\alpha = 9.75 \times 10^{-40} R^{6}_{6} /
I_{45}$ s, $R_{6} = R/ 10^{6}$cm , $I_{45} = I / 10^{45}$ g$\cdot$
cm$^{2}$, $I$ is the moment of inertia. We assume that the spin
and magnetic axes are perpendicular. The decay of the magnetic field 
decreases the spin-down rate
and, hence, changes the evolutionary tracks of pulsars both
in the B - P and B - $\tau$ planes.

\section{Numerical results}

We solve the equation (2) with the corresponding boundary and
initial conditions by making use of the implicit differencing scheme.
The original field is assumed to be confined to the outer layers
of the crust with the density $\rho \leq \rho_{0}$. The calculations
have been performed for a wide range of $\rho_{0}$, $5 \times 10^{13}$
g/cm$^{3} \geq \rho_{0} \geq 10^{10}$ g/cm$^{3}$. It was argued by
Urpin \& Muslimov (1992) that the decay is sensitive to 
the initial depth penetrated by the field and, hence, to the value
of $\rho_{0}$. However, the evolution is much less flexible to 
the particular form of the original field distribution within the
layer $\rho < \rho_{0}$. The only exception is the case when a
substantial fraction of the initial currents is mainly concentrated 
at $\rho \ll \rho_{0}$. In the present calculations, we choose 
$s(r,0)$ in the form
$$
s(r,0) = (1 - r^{2}/r_{0}^{2})/(1 - R^{2}/r_{0}^{2}) \;\; at \;\; 
r>r_{0}, \eqno(6)
$$
$$
s(r,0) = 0 \;\; at \;\; r<r_{0},
$$
where $r_{0}$ is the boundary radius of the region originally
occupied by the magnetic field, $\rho_{0} = \rho(r_{0})$.

Unfortunately, at present, there is no plausible estimate of
the impurity parameter, $Q$, in the neutron star crust. In our
calculations, the value $Q$ is taken within the range $0.1 \geq
Q \geq 0.001$, and it is assumed to be constant throughout the
crust.

To simulate the rotational evolution one requires the value of the
moment of inertia, $I$, which is determined by the equation of state. 
We use for $I$ the simple analytic expression suggested by Ravenhall
\& Pethick (1994) and acceptable for a wide variety of equations of
state. 

Fig.2 shows the evolution of the surface magnetic field, normalized
to its initial value, $B_{0}$. The top and bottom panels represent 
the models with standard and accelerated cooling, respectively.
It is seen that the decay is qualitatively different for
different cooling scenarios. For the slowly cooling standard models,
the higher crustal temperature leads to a lower electrical
conductivity and, hence, to a more rapid decrease of the field
during the neutrino cooling era. The rate of dissipation of the
magnetic field is particularly high during the early evolutionary
stage when the conductivity is determined by electron-phonon
scattering in the region in which currents are concentrated. During
this short stage ($\leq 1$ Myr), the surface field strength can
weaken by a factor of 5-1000, depending on the original depth 
penetrated by the field and the equation of state. The stiffer is
the equation of state, the slower is the field decay at a fixed 
$\rho_{0}$. This dependence is evident because the rate of 
dissipation is determined by the length scale of the field 
which is larger for the neutron star with a stiffer equation
of state. Note that the difference in the field strength
may be very large after the initial stage ($t \leq 1$ Myr)
for the models with different equations of state even if the
original field is confined to the layers with the same $\rho_{0}$.
Thus, the field confined originally to the density $\rho =
10^{13}$ g/cm$^{3}$ decreases by a factor of $\sim 5$, $33$
and $100$ after $1$ Myr for the PS, FP and BPS models,
respectively.

Between 0.1 and 2 Myr, the dominant conductivity mechanism 
changes from electron-phonon to electron-impurity scattering
in the standard cooling model. As a result, the conductivity
increases and the rate of field decay slows down. The
characteristic feature of the models with standard cooling
is the presence of flat portions of the decay curves at
$t \sim 1-300$ Myr depending on the impurity parameter. The
lower is the impurity content, the longer is the plateau on the 
corresponding decay curve. These plateaus reflect a change of
the conductivity regime in which $\sigma$ becomes relatively
high and independent of $T$. The length of plateau depends
on the equation of state at given $Q$. Thus, at $Q=0.001$,
the field is practically constant during $\sim 3 \times 10^{7}$,
$2 \times 10^{8}$, $3 \times 10^{9}$ yr after the initial stage
for the BPS, FP and PS models, respectively. During
the impurity dominating stage, the decay may be extremely
slow if the impurity content is low. Note that the decay follows 
approximately a power law during the 
late evolutionary stage when the neutron star leaves a plateau.
This simple dependence can be obtained from the analytical
consideration of diffusion of the magnetic field in the crust
(see Urpin, Chanmugam \& Sang 1994). A power law decay lasts until
the magnetic field reaches the crust-core boundary. After this
point, the decay becomes faster. Departures from the power law 
are especially pronounced for the BPS model after $10^{8}$ yr,
since the radius is smallest and the diffusion time-scale is shortest
for this model. For other models, diffusion throughout
the crust proceeds on a time scale longer than $10^{9}$ yr,
thus departures are negligible at $t \leq 10^{9}$ yr.

The accelerated cooling models result in substantially
different field decay. Most noticeable is the slow decay - depending
on the original depth and the equation of state, the field
weakens by a factor 1.5-100 after $10^{9}$ yr. This slow decay
is due to the higher conductivity in the cooler crust. The decay
curves for the accelerated cooling models do not practically
exhibit the plateaus as do the curves for the standard models.
The only exception is the case when the original field is 
confined to the layers with a small depth ($\rho_{0} \leq
10^{11}$ g/cm$^{3}$) but even for such magnetic configurations
the plateaus are much less pronounced than in the top panels.
This is due to the fact that the internal temperature falls
down very quickly for the neutron star with accelerated
cooling. In reality, impurity scattering dominates the 
conductivity in the deep layers practically from the 
beginning and those magnetic configurations which are
anchored in these layers do not experience a change of the
conductivity regime. The decay of such a deeply anchored
field is monotonic. However, inspite of a low temperature of the
accelerated cooling models, the conductivity can be mainly determined
by electron-phonon scattering in the layers with a relatively low 
density ($\rho \leq 10^{11}$ g/cm$^{3}$) during the initial
evolutionary stage. If the field is originally confined to
the layers with such a low density, the decay curves can exhibit
the plateaus but, of course, they lie much above the corresponding
plateaus for the standard cooling models since the crustal 
conductivity is much higher and the decay is much slower for the
rapidly cooling models.

Figure 3 shows the strength of the surface magnetic field at the 
pole versus the chatacteristic age, $\tau = P/ 2 \dot{P}$. For all
models, the calculations have been done for the initial polar
magnetic field $B_{o} = 3 \times 10^{13}$ G and the initial
period $P_{0} = 0.01$ s. Note that the behaviour of tracks in
the B-$\tau$ plane are sensitive to the initial field strength
since the spin evolution depends non-linearly on the magnetic
field. On the contrary, the decay curves in Fig.2 do not depend
on $B_{0}$ because the magnetic evolution is determined by
the linear equation (2). The dependences of $B$ on $\tau$ are
qualitatively similar to the decay curves: the standard cooling
models exhibit the plateaus whereas the accelerated cooling
models show a much more monotonic behaviour. The tracks are
weakly sensitive to the initial spin period $P_{0}$: the
neutron star forgets about its initial rotation
after a relatively short time (of course, if $P_{0}$ is not
originally very large). The magnetic evolution in terms of the spin-down
age $\tau$ proceeds slower than in a real time because one
always has $\tau(t) > t$ for a decreasing magnetic field. Thus,
the tracks for the BPS model with standard cooling reach the
plateaus after $\tau \sim 3-100$ Myr depending on the initial
depth penetrated by the field. These spin-down ages correspond
to $t \sim 1$ Myr. Obviously, the difference is less pronounced
for the FP and PS models which experience a slower decay.

\section{Discussion} 

A comparison of the computed tracks with the available observational
data on pulsar magnetic field allows, in principle, to infer some
parameters of their magnetic configurations. Note, however, that 
the interpretation of pulsar data is a sublect of many uncertainties,
and there is no commonly accepted point of view which behaviour 
of the magnetic field is most suitable to account for a great 
variety of observations. 

A some evidence of a relatively slow field decay has been obtained 
recently by making use of the so called method of population
synthesis (Bhattacharya et al. 1992, Wakatsuki et al. 1992, 
Hartman et al. 1996). In this
method, one assumes distributions for the initial pulsar parameters,
and laws governing their evolution, to simulate a population of
radio pulsars. The main advantage of the population synthesis method
is to model the selection effects in detail (Lorimer et al. 1993).
Comparing the simulated population with the observed one, Bhattacharya
et al. (1992) and Hartman et al. (1996) concluded that models in 
which the magnetic field decays little during the active life time
of a radio pulsar give the best description of the observations. This
conclusion seems to be weakly sensitive to many assumptions concerning
the initial pulsar properties and their evolution. For example,
Hartman et al. (1996) used the improved information that has become
available since the paper by Bhattacharya et al. (1992) was written,
on the distribution of electrons, on the velocity 
distribution of newly born neutron stars and on the effect of
gradient in the birthrate in the Galaxy. In spite of a large difference
in these input parameters, the authors of the both papers concluded
that the models with long decay times ($\geq 30$ Myr) give acceptable 
fits to the observational data. According to Hartman et al. (1996), 
good fits can be obtained if the mean magnetic field of the pulsar 
population ranges from $2 \times 10^{12}$ to $4 \times 10^{12}$ G. 
Note that these conclusions has been obtained for the particular case 
of the exponential field decay which appears to be rather questionable 
from the theoretical point of view. Besides, the exponential decay 
leads to a very fast decrease of the field for ages longer than 
the decay time-scale. Due to this, the exponential law can be
satisfactory for making old pulsars with relatively weak magnetic fields 
($\sim 10^{11}$ G) in sufficient numbers only if the decay time is
comparable with the true age of oldest pulsars.  The situation may be 
quite different for other decay laws which are characterized by a slower 
decrease than the exponent. Note also that the recent analyses of slowly
rotating pulsars (see Han 1997) indicates some problems in 
the evolution of these objects if the fields decay exponentially.

Probably, one of disadvantages of the population synthesis method
concerns very young pulsars with $t \leq 10^{5}$ yr. The number of these 
pulsars is small and, correspondingly, they give a negligible 
contribution to any statistical analysis. The behaviour of newly
born pulsars during first $10^{5}$ yr of their life, however, may be
very important for our understanding of the neutron star physics.
It has been argued by Lyne (1994) that the magnetic fields of pulsars
in supernova remnants are appreciably stronger than the average 
field of the main population. The number of these youngest known
pulsars is relatively small and does not allow to make a statistically
reliable conclusion at present. However, if this point is correct
then either pulsars in supernova remnants are not progenitors of the
standard pulsar population or the magnetic field has to decay. Note 
once more an impotance of selection effects for any analysis of the 
pulsar population. In the case of supernova remnants, these effects
favour pulsars with high magnetic fields because they are bright and 
can be detected agains the strong emission from the shell and plerion. 
Also, many pulsars in supernova remnants are discovered accidentally
rather than systematically, and many could have been missed.

In Fig.4, we plot the magnetic field versus the spin-down age $\tau$
for 440 pulsars taken from the paper by Taylor, Manchester \& Lyne
(1993). These data have been completed by the most recent data on 
pulsars in supernova remnants published by Frail, Goss \& Whiteoak 
(1994). We did not plot in this figure binary pulsars, pulsars 
in globular clusters and those pulsars which has the magnetic  
field $< 10^{10}$ G. Note that almost all these low-magnetized 
pulsars (with the exception of two) either enter binary systems or
have a period shorter than 100 ms and may be related to millisecond
pulsars. The magnetic and spin evolution of all excluded objects 
may be essentially influenced by mass transfer and may differ from 
that of isolated pulsars. Pulsars in supernova remnants 
are marked by starlets. We calculated the polar strength of the magnetic
field from $P$ and $\dot{P}$ assuming the FP equation of state.
Fig.4 shows a clear decrease in the surface field strength with 
increasing $\tau$. However, as pointed out by Lyne, Ritchings \& 
Smith (1975), an appreciable portion of the above trend may be caused
by a much larger range in $\dot{P}$ than in $P$ of the observed pulsars, 
thus the $B-\tau$ distribution cannot be a reliable argument for
the field decay. Nevertheless, the characteristic  
magnetic field of very young pulsars can be well inferred from this
figure. A tendency of youngest pulsars to have the magnetic field
$\sim (1-2) \times 10^{13}$ G is clearly seen from the plot. This
value is much higher than the average field required for good fits
in population synthesis method.

In our mind, both the above points can be easily understood in the
frame of the crustal magnetic field model. The decay scenarios for
neutron star models with standard cooling give a good
compromise between the results of Lyne (1994), on one hand, and
Bhattacharya et al. (1992) and Hartman et al. (1996), on the other
hand. Actually, the decay curves for standard cooling show a 
relatively rapid decrease of the magnetic field at the beginning
of the evolution, at $t \leq 10^{5}$ yr (see Fig.2), for any
equation of state. During this short initial phase, the field strength
can be reduced from the value $\sim (6-30) \times 10^{12}$ G, typical
for pulsars in supernova remnants, to the "standard" pulsar field
$\sim (1-3) \times 10^{12}$ G. Pulsars with such short ages will 
not evidently contribute to the statistical properties of the pulsar
population because of a small number of such young objects. After
the initial phase, at $t > 10^{5}$ yr, the decay of the crustal field 
slows down and the field can be practically unchanged during a long 
time. The age of the great majority of known pulsars lies within the
time interval when decay curves show the plateaus, thus, our theory
provides a natural scenario of a slow decay for these pulsars. When
comparing the simulated population obtained in population synthesis
with the observed one, these pulsars give the dominating contribution.

The parameters most suitable for the initial magnetic configurations
of neutron stars and their crust can be easily estimated if one 
adopts the above scenario. As it was mentioned, the magnetic field
of young pulsars in supernova remnants is usually by a factor 3-10
higher than the average pulsar field. Since the field strength on the
plateaus in Fig.2 is determined by the initial depth penetrated
by the field, one can estimate $\rho_{0}$ which corresponds to such
a decrease during the initial phase. This density is approximately
$(1.5-2) \times 10^{14}$, $(5-10) \times 10^{13}$ and $(1-10) \times
10^{12}$ g/cm$^{3}$ for the BPS, FP and PS models, respectively.
The initial field should occupy practically the whole crust in the case
of the BPS model, however, the fraction of the occupied crust volume 
is much smaller for the FP and PS models. The corresponding depth 
ranges from $\approx 550$ to $\approx 650$ m for the FP model and
from $\approx 900$ to $\approx 1100$ m for the PS model.

The length of plateaus on the decay curves (see Fig.2) is determined
by the impurity parameter, $Q$. The population synthesis method gives 
acceptable fits for observations if the magnetic field decays little 
on a time scale comparable with the active life time of a radio pulsar,
$100 \geq t \geq 30$ Myr (see Bhattacharya et al. 1992, Hartman et al.
1996). Practically
for all considered equations of state, the plateaus go until 
$t \sim 30-100$ Myr if $Q \approx 0.01-0.03$. If the decay time 
acceptable for good fits will turn out to be shorter (for example, for
other decay laws), then the impurity parameter may be even larger.
Unfortunately, at present there is no plausible ideas regarding the 
impurity content in the neutron star crust, however, the value
$Q \sim 0.01$ seems to be consistent with the generally used estimate
for this parameter, $Q \sim 0.1-0.001$, and corresponds to an
intermediately polluted crust.

The models with accelerated cooling seem to be less attractive for
an interpretation of the above data on pulsar magnetic fields. The 
decay during the initial phase is much less pronounced for these
models and the initial phase is longer than in the case of standard
cooling.

In Fig.5 we show the evolutionary tracks in the $B-P$ plane for
different initial conditions and equations of state assuming the
standard cooling scenario. We also plot in this figure the observed
distribution of radio pulsars. Note that the observed positions of 
radio pulsars depends on the equation of state since the magnetic
field is calculated from the data on $P$ and $\dot{P}$ by making
use of equation (5) which depends on the moment of inertia. The
latter, in its turn, is determined by the equation of state. This
is the reason why the measured magnetic fields are different for 
the BPS, FP and PS models in Fig.5. We choose the parameters of 
tracks with the only wish to illustrate how the pulsars with 
different original magnetic configurations move in the $B-P$ plane 
and to show that the observed variety of isolated
pulsars with a wide range of measured magnetic fields and periods 
can be naturally formed in the frame of our model. It turns out,
however, that the hypothesis of the crustal magnetic field is not
satisfactory for the neutron star models based on soft equations of
state. For the BPS model, it is rather difficult to account for
the existance of pulsars with a high magnetic field, $B \sim (3-5)
\times 10^{13}$ G, and with a long period, $P \sim (2-5)$ s for
any reasonable choice of the initial parameters. Such pulsars cannot
be formed in the course of evolution even if the original magnetic
field occupies the whole crust ($\rho_{0} = 2 \times 10^{14}$ 
g/cm$^{3}$) and is as high as $10^{14}$ G. The main reason of this 
is small radius and thickness of the crust of soft models and, hence,
a relatively fast decay of the crustal magnetic field. Due to a fast
decay, the energy loss for magnetodipole radiation decreases very 
rapidly and the neutron star cannot slow down to long periods
$\sim 2-5$ s while having a high magnetic field. This problem
concerns all models with the equation of state softer than that
of Friedman-Pandharipande.

For stiffer equations of state, the situation is more optimistic
and the observational data can be well accounted for the hypothesis
of the crustal origin of the magnetic field. The radius and thickness
of the crust is larger for stiffer models and, therefore, the field
decay is much slower. In contrast to a widely accepted opinion
that any model with the crustal magnetic field should have troubles
in the explanation of the evolutionary history of relatively old 
pulsars with strong magnetic fields, it turns out that models with 
stiff equations of state are quite suitable for this. Even the 
the existence of the oldest among highly magnetized pulsars can be 
easily understood in the frame of the crustal model. This can be well
illustrated by the curve 3 in the PS(s) panel of Fig.2. This decay 
curve corresponds to the magnetic configuration which occupies 
initially only $\sim 1/4$ of the thickness of the crust, however,
the decay is already extremely slow in this case: the field is 
reduced by a factor $\sim 10$ after $10^{10}$ yr. Therefore, there 
is nothing particular for the neutron star to have the magnetic 
field $\sim 10^{12}$ after $10^{10}$ yr of the evolution if the 
initial field strength was more or less standard, $\sim 10^{13}$ G.
For instance, the famous pulsar in Her X-1 which has the magnetic 
field of the order of $(1-3) \times 10^{12}$ G and is about 600 Myr 
old (Verbunt, Wijers \& Burm 1990) requires even a less strong
initial field, or its initial field can be confined to less dense 
layers.

Note that all considered models have no troubles in making 
a sufficient number of weakly magnetized and shortly periodic isolated 
pulsars. Such pulsars can be formed if either the initial depth 
penetrated by the field is smaller than for the mean pulsars, or the
parameter $Q$ is a bit larger, or the initial field strength is
weaker (see Fig.5).

The parameters of the
initial magnetic configurations required for a description of observed
pulsars are more or less standard. For example, the initial strength
of the magnetic field can probably lie within the range $3 \times
10^{13} \geq B_{0} \geq 10^{12}$ G for the majority of pulsars if
the equation of state is close to that of Pandharipande-Smith. The
initial depth penetrated by the magnetic field ranges typically from 
$10^{12}$ to $10^{13}$ g/cm$^{3}$ for the main pulsar population
and only very few pulsars requires the original field anchored in
bit deeper layers. This conclusion seems to be in a good agreement
with the estimate of $\rho_{0}$ obtained above. Obviously, the
original field has to be anchored in more deep layers for models
with softer equations of state. The FP model, for instance, gives
an acceptable description of the pulsar population if the field 
occupies originally the crustal layers with $\rho_{0} \sim 10^{13}-
10^{14}$ g/cm$^{3}$. Correspondingly, the initial field should
be stronger for this model, $6 \times 10^{13} \geq B_{0} \geq 3 \times
10^{12}$ G. The value of the impurity parameter $Q \sim 0.1-0.01$ 
appears to be acceptable for the both stiff and intermediate models.

\section{Conclusion}

We have examined the influence of the equation of state and the
cooling scenario on the decay of magnetic field anchored in the 
neutron star crust. The field decay depends on the radius of the
star and the thickness of its crust and, hence, is sensitive to the
equation of state of superdense matter. Besides, the
decay is also influenced by the physical processes in the neutron
star core and is qualitatively different for the different
cooling scenarios. The spin-down torque on the isolated pulsar is 
usually associated with the energy loss for magnetodipole radiation
and, therefore, depends on the behaviour of the magnetic field. 
In principle, these facts allow to infer some 
properties of superdense matter from analysis of the magnetic and spin 
evolution of neutron stars.

A stiffness of the equation of state determines the thickness of the 
crust as well as the moment of inertia of the neutron star. The stiffer 
is the equation of state,  the thicker is the crust and the larger is
the moment of inertia. The rate of the field decay depends on the 
length-scale
of the magnetic field, therefore, the neutron star with a softer
equation of state experiences a faster decay. The spin-down rate is
inversely proportional to the moment of inertia at fixed $P$ and $B$
(see equation (5)). Thus, both the magnetic and spin evolutions proceed 
slower for the neutron star model with a stiffer equation of state.
A difference can be rather appreciable for the considered neutron
star models. For example, the strength of the magnetic field is weaken 
after $10^{8}$ yr by a factor $\sim$300 for the BPS model and only by a 
factor $\sim$5 for the PS model if the field is originally confined to 
the layers with $\rho_{0} = 10^{13}$ g/cm$^{3}$ for both models and 
if the impurity parameter $Q=0.01$. 

Due to small radius and crust thickness, the magnetic field diffuses 
throughout the crust much faster for the BPS model than for the FP and 
PS models. Thus, at $Q=0.01$, the field reaches the bottom of the crust 
after $\sim 10^{8}$ yr for the BPS model practically independently of the 
initial distribution. The corresponding diffusion time-scale for the FP and 
PS models is longer than $10^{9}$ yr. A short diffusion time-scale
is the reason of a difference in the long-term behaviour of the
magnetic field for the BPS model in comparison with the FP and PS models.
It has been argued by Urpin et al. (1994) that, in cold neutron stars,
the surface strength of the magnetic field follows approximately
the power law dependence until the field reaches the crust-core
boundary. Since the diffusion time-scale is relatively short for the
BPS model, the power law decay stage is also short. After the point
when the magnetic field reach the superconductive core, the decay becomes
exponential. For the FP and PS models, a power law decay lasts much
longer (of course, if the crust is not strongly poluted by impurities). 

The magnetic evolution turns out to be sensitive to the cooling
scenario. For slowly cooling neutron stars with the standard neutrino
emissivity, the magnetic field decay is particularly fast during
the early evolutionary stage when electron-phonon scattering dominates
the crustal conductivity. After this stage ($t \leq 1$ Myr),
the surface field of the neutron star can be weaken by a factor 5-1000
depending on the initial depth penetrated by the field. Evidently,
the field decreases stronger for the model with a softer equation of state.
However, the decay slows down drastically at $t \geq 1$ Myr
because the dominant conductivity mechanism changes from 
electron-phonon to electron-impurity scattering and the conductivity 
increases after this initial phase. The characteristic point of the
models with standard cooling is the presence of plateaus on the
decay curves. The length of these plateaus depends on the impurity
parameter $Q$ and may vary from very extended for a low impurity content 
practically to zero for a highly poluted crust. During the impurity 
dominating stage, the decay follows the power law dependence 
until the magnetic field approaches the boundary between the 
crust and superconductive core. This stage may be very long ($\geq 
10^{9}$ yr) for the FP and PS models but is much shorter for the BPS 
model. 

The accelerated cooling models show a qualitatively different behaviour.
In this case, the temperature falls down drastically soon after the 
neutron star is born, therefore, the crustal conductivity is higher 
for these models during the whole evolution. A higher conductivity leads 
to a substantially slower decay. Thus, depending on the equation of
state, the field can be weaken only by a factor 1.5 - 100 after $10^{9}$
yr if the field is originally confined to the layers with $\rho_{0}$
within the range $10^{11}-10^{13}$ g/cm$^{3}$. The decay curves for
the accelerated cooling models do not practically exhibit the plateaus
as do the curves for the slowly cooling models. The reason for this 
is that impurity scattering dominates the conductivity of deep layers
practically from the beginning of evolution, and those magnetic 
configurations which are anchored in these layers do not experience
the change of conductivity regime. 
	 
The magnetic and rotational evolution of observed pulsars may be
easily understood if their magnetic fields are of the crustal
origin. The models with standard cooling seem to be most suitable 
for a description of observational data on the pulsar magnetic fields
due to the fact that these models are characterized by a fast decay
during the short initial phase and a very slow decay in the course 
of the subsequent evolution. This evolutionary behaviour allows
to account for in a natural way why the field of young pulsars in
supernova remnants is typically stronger than that of the main
pulsar population (see Lyne 1994) and why the decay is very slow
for not very young pulsars (see Bhattacharya et al. 1992, Hartman
et al. 1996). The neutron star models with accelerated cooling
don't show such a behaviour and, in this sense, they are less
consistent with the available data on the pulsar magnetic fields.

The magnetorotational evolution of neutron stars with the crustal
magnetic field is in a good agreement with the observed distribution
of pulsars in the $B-P$ plane only if the equation of state of nuclear
matter is sufficiently stiff. The evolution of neutron star models with
equations of state softer than that of Friedman-Pandharipande is likely
inconsistent with the $B-P$ distribution of radio pulsars because the
radius and crustal thickness are small for these models and, hence, 
the field decay is relatively fast. Therefore, it is difficult to
explain the existence of pulsars with high magnetic fields ($\sim
(2-5) \times 10^{13}$ G) and long periods ($\sim (2-5)$ s) in the
frame of such models. This problem does not arise if the equation of 
state of nuclear matter is stiffer. In our mind, the models within  
the range from FP to PS equations of state are most suitable for
a description of the $B-P$ data.

Note in conclusion that the model of neutron stars with the crustal
magnetic field does not require extraordinary values of parameters of 
both the initial magnetic configuration and the crust in order to be 
consistent with observations. An acceptable agreement, for instance,
for the PS model can be obtained if the initial field strength is
more or less standard, $B_{0} \sim 10^{12} - 3 \times 10^{13}$ G,
and the depth penetrated originally by the field is confined to the
density $\rho_{0} \sim 10^{12} - 10^{13}$ g/cm$^{3}$.

\section*{Acknowledgement}

The authors thank the referee, Dr R.A.M.J.Wijers, for constructive
suggestions concerning the interpretation of observational data.
A financial support is gratefully acknowledged under the
Grant 94-02-06540(a) of the Russian Foundation of Basic Research and
94-3097 Grant of INTAS.

\section*{References}

Baym G., Pethick C.J., Sutherland P.G., 1971, ApJ, 170, 299 \\
Bhattacharya D., Wijers R., Hartman J., Verbunt F., 1992, A\&A, 254, 198 \\
Frail D., Goss W., Whiteoak J., 1994, ApJ, 437, 781 \\
Friedman B., Pandharipande V.R., 1981, Nucl. Phys. A, 361, 502 \\
Han J.L., 1997, A\&A, 318, 485 \\
Hartman J., Bhattacharya D., Wijers R., Verbunt F., 1996, A\&A (in press) \\
Itoh N., Hayashi H., Kohyama Y., 1993, ApJ, 418, 405\\
Lattimer J., Pethick C., Prakash M., Haensel P. 1991, Phys.Rev.Lett.,                           
66, 2701 \\
Lorimer D., Bailes M., Dewey R., Harrison P., 1993, MNRAS, 263, 403 \\
Lyne A., Ritchings R., Smith F., 1975, MNRAS, 171, 579 \\
Lyne A., 1994, in "Lives of Neutron Stars", eds. A.Alpar, \"U.Kiziloglu
\& J.van Paradijs (Dordrecht: Kluwer), 213 \\
Nomoto K. \& Tsuruta S. 1987, ApJ, 312, 711 \\
\"Ogelman H., 1994, in "The Lives of the Neutron Stars", eds. A.Alpar,
\"U.Kiziloglu \& J.van Paradijs (Dordrecht: Kluwer), 101 \\
Ostriker J.P., Gunn J.E. 1969, ApJ, 157, 1395 \\
Pandharipande V.R., Pines D., Smith R.A., 1976, ApJ, 208, 550 \\
Ravenhall D., Pethick C. 1994, ApJ, 424, 846 \\
Sang Y., Chanmugam G. 1987, ApJ, 323, L61 \\
Sang Y., Chanmugam G. 1990, ApJ, 363, 597 \\
Taylor J., Manchester R., Lyne A., 1993, ApJS, 88, 529 \\
Thompson C., Duncan R. 1993, ApJ, 408, 194 \\
Urpin V., Chanmugam G., Sang Y. 1994, ApJ, 433, 780 \\
Urpin V., Levshakov S., Yakovlev D. 1986, MNRAS, 219, 703 \\
Urpin V., Muslimov A. 1992, MNRAS, 256, 261 \\
Urpin V., Van Riper K. 1993, ApJ, 411, L87 \\
Van Riper K.A. 1991, ApJS, 75, 449 \\
Verbunt F., Wijers R.A.M.J., Burn H.M.G. 1990, A\&A, 234, 195 \\ 
Wakatsuki S., Hikita A., Sato N., Itoh N. 1992, ApJ, 392, 628 \\
Yakovlev D., Urpin V. 1980, SvA, 24, 303 \\                                                         

\newpage

\begin{figure}[p]
\vspace{15cm}
\input epsf
\epsfbox[90 150 10 30]{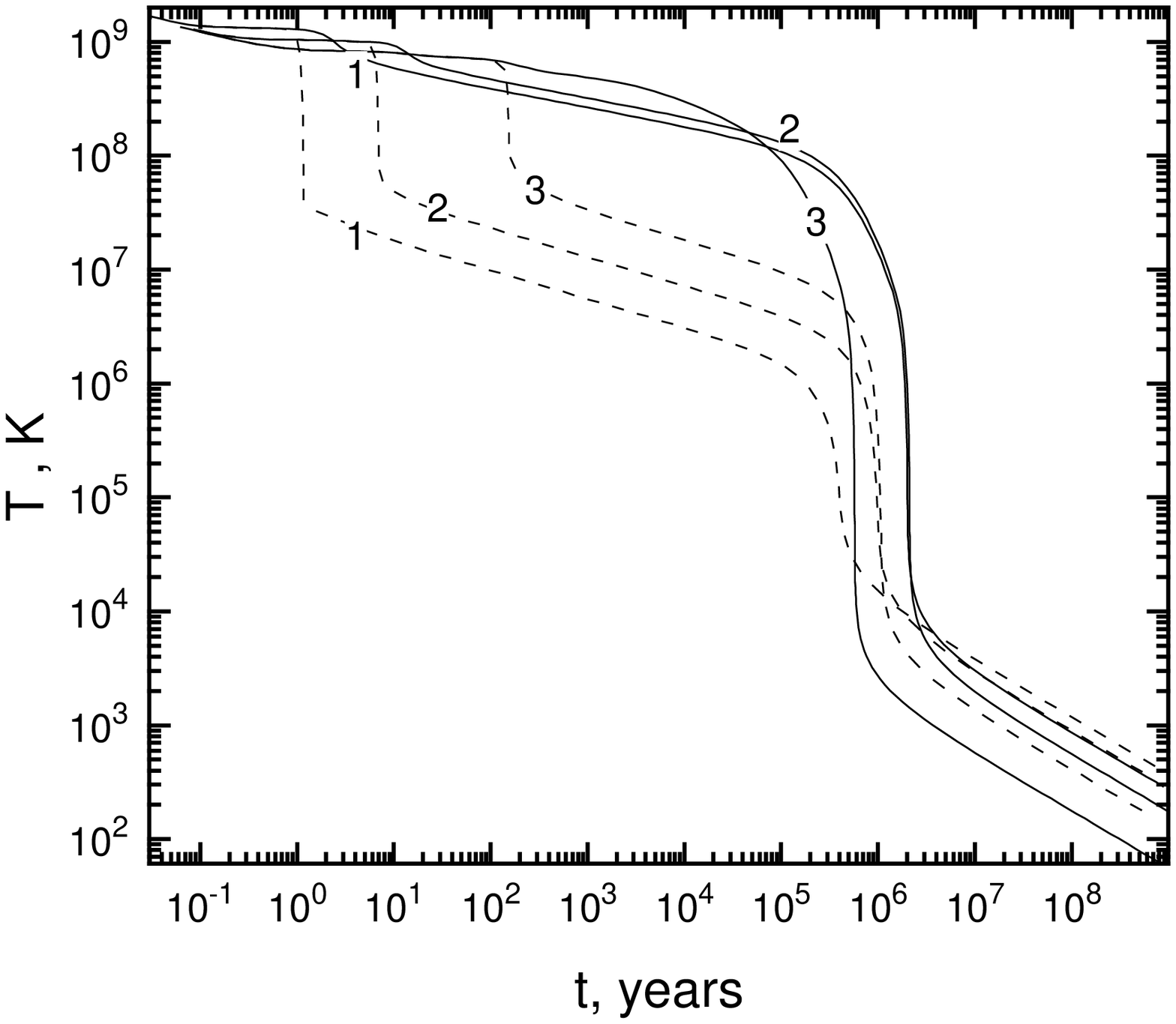}

\caption{The dependence of the internal temperature, $T$, on
the neutron star age for different equations of state and cooling 
scenarios. Solid and dashed lines correspond to the models with
standard and accelerated cooling, respectively. Numbers near the
curves indicate the BPS (1), FP (2) and PS (3) equations of state.}
\end{figure}

\clearpage
\begin{figure}[t]
\vspace{20cm}
\input epsf
\epsfbox[90 190 10 30]{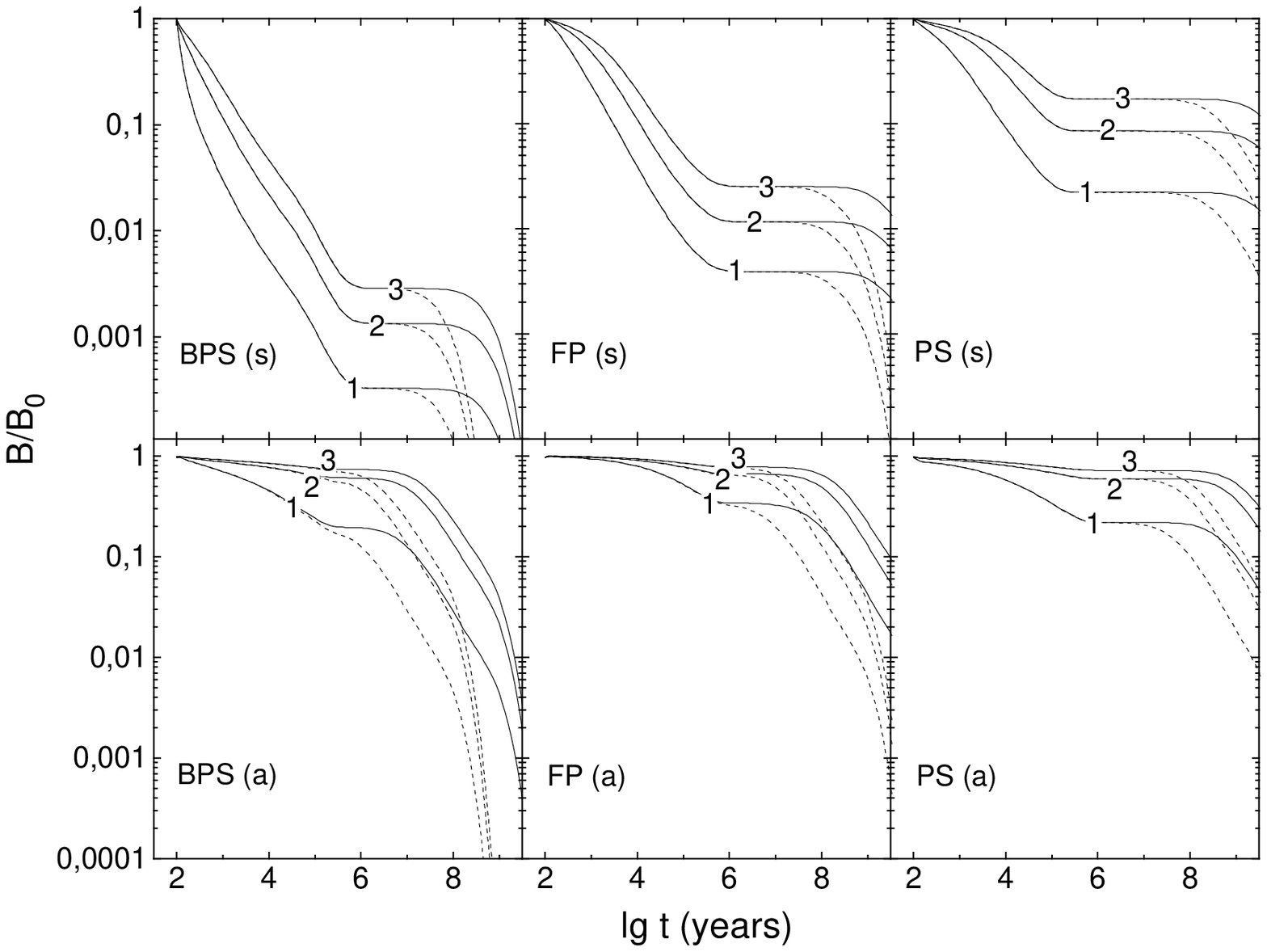}

\caption{The evolution of the magnetic field normalized to its
initial value for the neutron star models with the BPS, FP and
PS equations of state and with standard (s) and accelerated (a) 
cooling. The numbers near the curves correspond to different
values of the initial depth penetrated by the field; $\rho_{0} = 
10^{11}$ (curve 1), $10^{12}$ (2) and $10^{13}$ (3) 
g/cm$^{3}$. The solid lines show the decay for $Q=0.001$, the
dashed lines - for $Q=0.01$.}
\end{figure}

\clearpage
\begin{figure}[t]
\vspace{20cm}
\input epsf
\epsfbox[90 190 10 30]{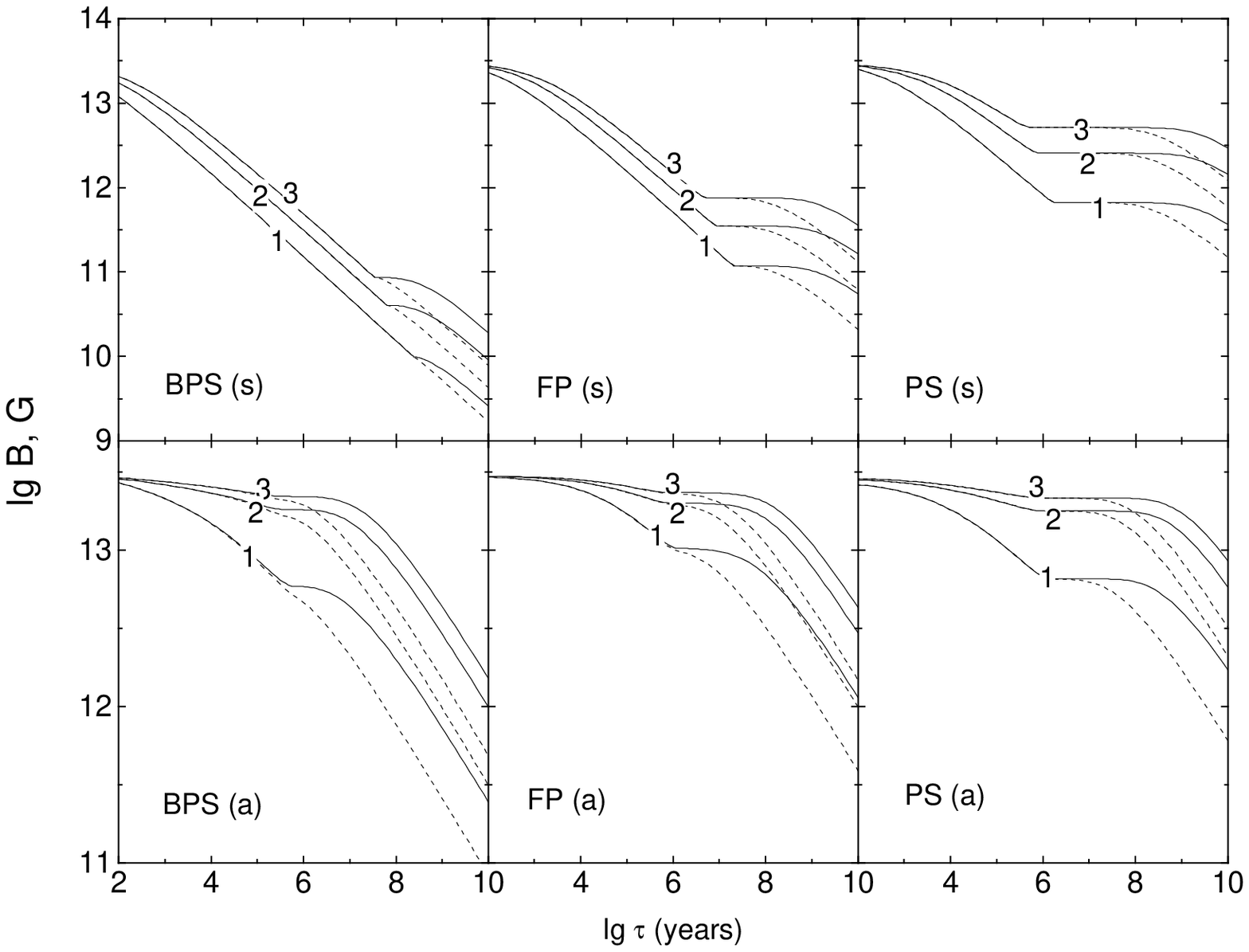}

\caption{The dependence of the magnetic field on the spin-down
age, $\tau = P/2 \dot{P}$, for the neutron star models with the BPS, 
FP and PS equations of state and with standard (s) and accelerated (a)
cooling. The numbers near the curves correspond to $\rho_{0} =
10^{11}$ (curve 1), $10^{12}$ (2) and $10^{13}$ (3) g/cm$^{3}$. The
solid lines show the decay for $Q=0.001$, the dashed lines - for
$Q=0.01$.}
\end{figure}

\clearpage
\begin{figure}[t]
\vspace{15cm}
\input epsf
\epsfbox[110 110 10 30]{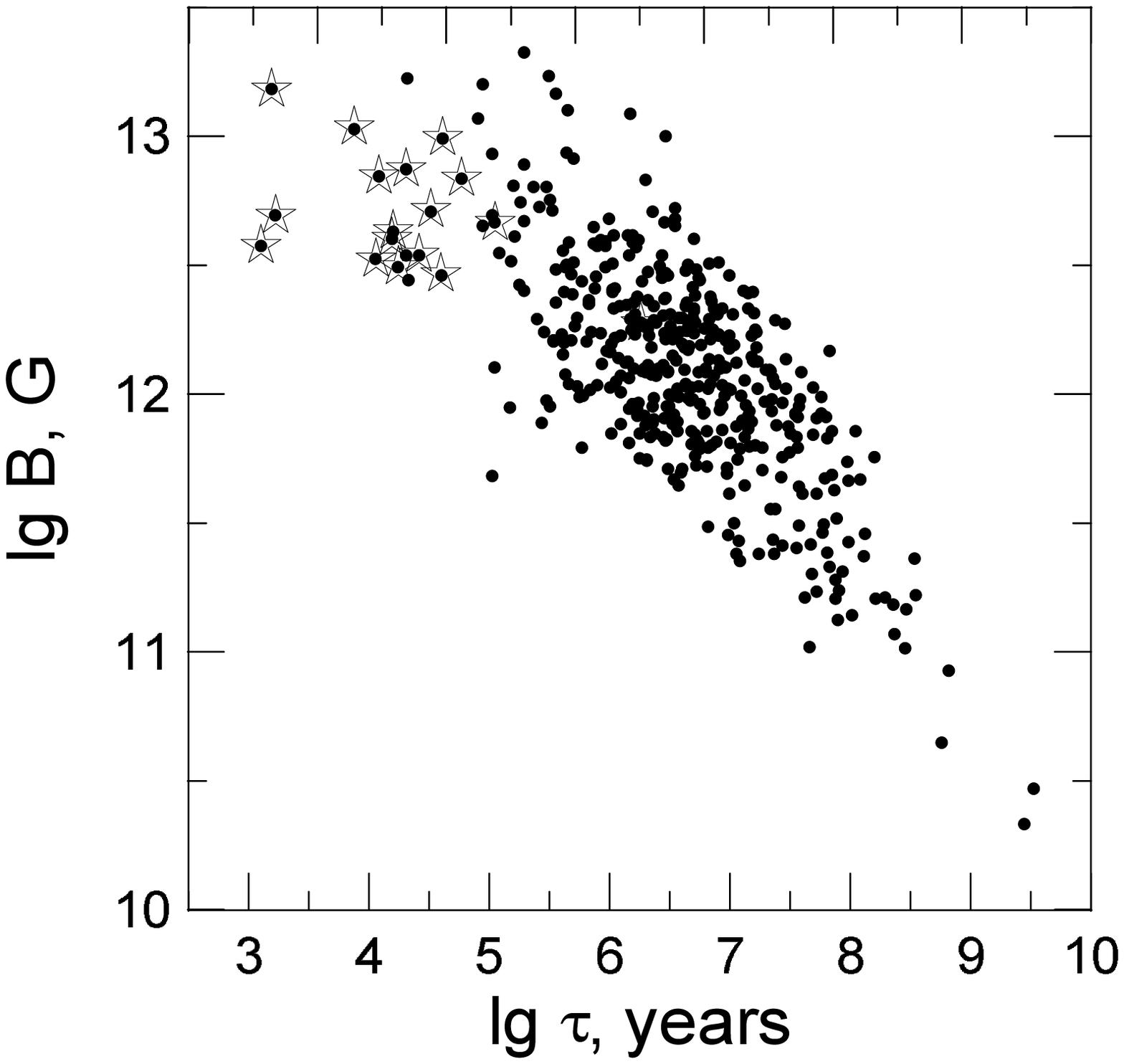}

\caption{The magnetic field, $B$, versus the spin-down age,
$\tau = P/2 \dot{P}$, for 440 observed radiopulsars. The surface
field strength is calculated assuming the FP equation of state.
The pulsars in supernova remnants are marked by starlets.}
\end{figure}

\clearpage
\begin{figure}[t]
\vspace{10cm}
\input epsf
\epsfbox[70 150 10 30]{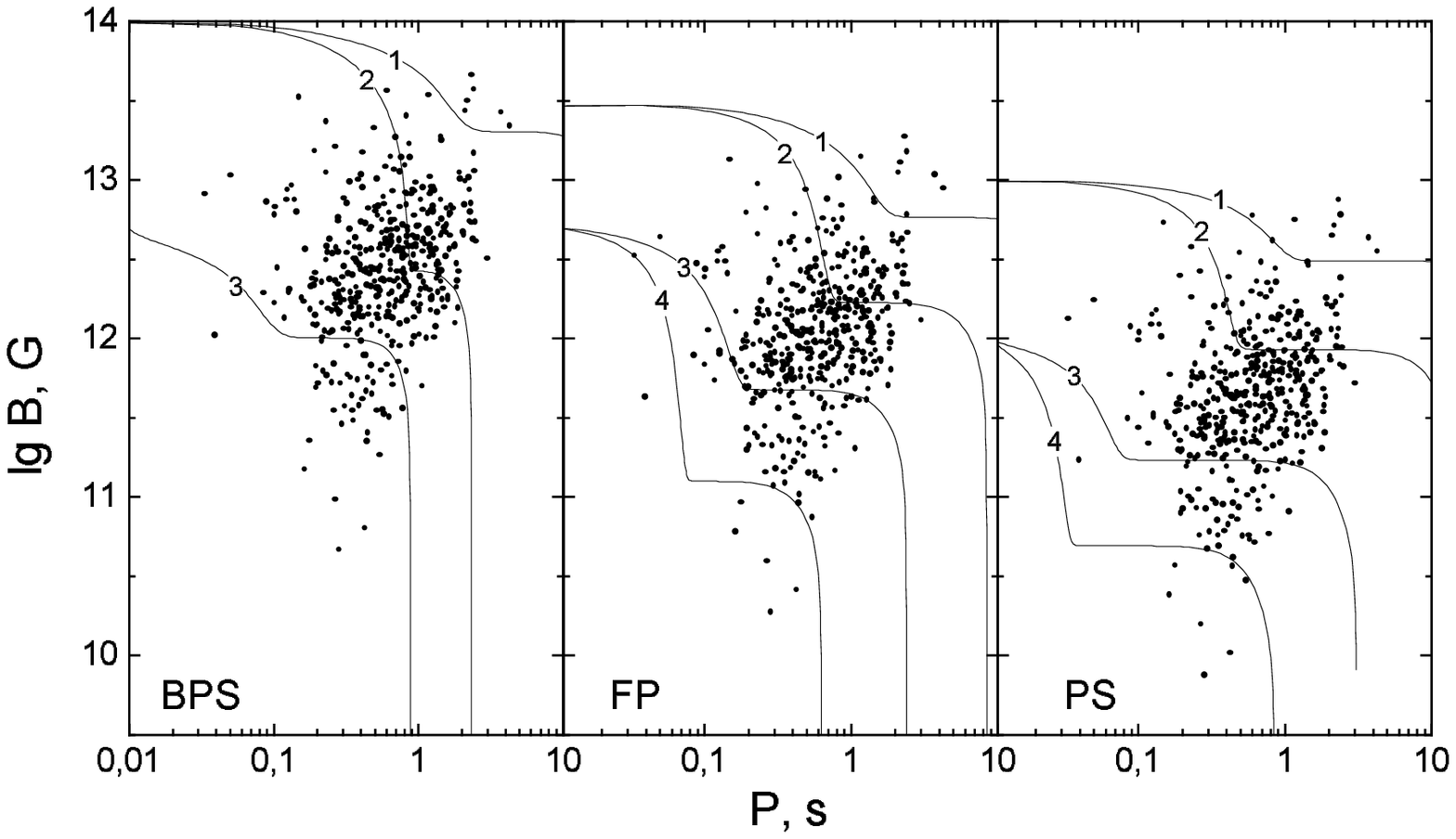}

\caption{The observable distribution of pulsars and the evolutionary
tracks in the $B-P$ plane for BPS, FP and PS models. The tracks 
correspond to different initial magnetic fields and depths penetrated 
by the field; in the BPS panel: $\rho_{0} = 2 \times 10^{14}$ (curve 1),
$10^{14}$ (2), $2 \times 10^{14}$ (3); in the FP panel: $\rho_{0} =
10^{14}$ (1), $3 \times 10^{13}$ (2), $5 \times 10^{13}$ (3), $10^{13}$
(4); in the PS panel: $\rho_{0} = 3 \times 10^{13}$ (1), $10^{12}$ (2),
$10^{13}$ (3), $3 \times 10^{11}$ (4) g/cm$^{3}$. All tracks are
calculated assuming $Q=0.01$.}
\end{figure}

\end{document}